\documentclass[prb,twocolumn,a4paper]{revtex4}
\usepackage{graphicx}
\usepackage{amsmath}

\begin{document}

\title{Coplanar Waveguide Resonators for Circuit Quantum Electrodynamics}
\author{M.~G\"oppl}
\affiliation{Department of Physics, ETH Z\"urich, CH-8093, Z\"urich, Switzerland.}
\author{A.~Fragner}
\affiliation{Department of Physics, ETH Z\"urich, CH-8093, Z\"urich, Switzerland.}
\author{M.~Baur}
\affiliation{Department of Physics, ETH Z\"urich, CH-8093, Z\"urich, Switzerland.}
\author{R.~Bianchetti}
\affiliation{Department of Physics, ETH Z\"urich, CH-8093, Z\"urich, Switzerland.}
\author{S.~Filipp}
\affiliation{Department of Physics, ETH Z\"urich, CH-8093, Z\"urich, Switzerland.}
\author{J.~M.~Fink}
\affiliation{Department of Physics, ETH Z\"urich, CH-8093, Z\"urich, Switzerland.}
\author{P.~J.~Leek}
\affiliation{Department of Physics, ETH Z\"urich, CH-8093, Z\"urich, Switzerland.}
\author{G.~Puebla}
\affiliation{Department of Physics, ETH Z\"urich, CH-8093, Z\"urich, Switzerland.}
\author{L.~Steffen}
\affiliation{Department of Physics, ETH Z\"urich, CH-8093, Z\"urich, Switzerland.}
\author{A.~Wallraff}
\affiliation{Department of Physics, ETH Z\"urich, CH-8093, Z\"urich, Switzerland.}
\date{\today}

\begin{abstract}
We have designed and fabricated superconducting coplanar waveguide
resonators with fundamental frequencies from $2$ to $9 \, \rm{GHz}$
and loaded quality factors ranging from a few hundreds to a several
hundred thousands reached at temperatures of $20 \, \rm{mK}$. The
loaded quality factors are controlled by appropriately designed
input and output coupling capacitors. The measured transmission
spectra are analyzed using both a lumped element model and a
distributed element transmission matrix method. The experimentally
determined resonance frequencies, quality factors and insertion
losses are fully and consistently characterized by the two models
for all measured devices. Such resonators find prominent
applications in quantum optics and quantum information processing
with superconducting electronic circuits and in single photon
detectors and parametric amplifiers.
\end{abstract}

\maketitle

\section{Introduction}

Superconducting coplanar waveguide (CPW) resonators find a wide
range of applications as radiation detectors in the optical, UV and X-ray frequency
range \cite{Mazin2002, Day2003, Zmuidzinas2004, Mazin2008, Vardulakis2008},
in parametric amplifiers \cite{Tholen2007, Lehnert2007, Bergeal2008},
for magnetic field tunable resonators \cite{Lehnert2007, Palacios-Laloy2007, Sandberg2008}
and in quantum information and quantum optics experiments
\cite{Wallraff2004a, Wallraff2005,
Schuster2005, Wallraff2007, Schuster2007a, Schuster2007b, Majer2007, Houck2007, Astafiev2007, Sillanpaa2007, Hofheinz2008, Fink2008}.

In this paper we discuss the use of CPWs in the context of quantum
optics and quantum information processing. In the recent past it has
been experimentally demonstrated that a single microwave photon
stored in a high quality CPW resonator can be coherently coupled to
a superconducting quantum two-level system \cite{Wallraff2004a}. This
possibility has lead to a wide range of novel quantum optics
experiments realized in an architecture now
known as circuit quantum electrodynamics (QED) \cite{Wallraff2004a}. The
circuit QED architecture is also successfully employed in quantum
information processing \cite{Blais2004} for coherent single qubit
control \cite{Wallraff2004a}, for dispersive qubit read-out \cite{Wallraff2005} and for
coupling individual qubits to each other using the resonator as a
quantum bus \cite{Majer2007, Sillanpaa2007}.

Coplanar waveguide resonators have a number of advantageous
properties with respect to applications in circuit QED. CPWs can
easily be designed to operate at frequencies up to $10 \, \rm{GHz}$
or higher. Their distributed element construction avoids
uncontrolled stray inductances and capacitances allowing for better
microwave properties than lumped element resonators. In comparison
to other distributed element resonators, such as those based on
microstrip lines, the impedance of CPWs can be controlled at
different lateral size scales from millimeters down to micrometers
not significantly constrained by substrate properties. Their
potentially small lateral dimensions allow to realize resonators
with extremely large vacuum fields due to electromagnetic zero-point fluctuations
\cite{Schoelkopf2008}, a key
ingredient for realizing strong coupling between photons and qubits
in the circuit QED architecture. Moreover, CPW resonators with large
internal quality factors of typically several hundred thousands can now be routinely realized
\cite{Frunzio2005, Baselmans2005, Barends2007, Aaron2008}.

In this paper we demonstrate that we are able to design, fabricate
and characterize CPW resonators with well defined resonance
frequency and coupled quality factors. The resonance frequency is
controlled by the resonator length and its loaded quality factor is
controlled by its capacitive coupling to input and output
transmission lines. Strongly coupled (overcoupled) resonators with
accordingly low quality factors are ideal for performing fast
measurements of the state of a qubit integrated into the resonator
\cite{Wallraff2005, Schreier2008}. On the other hand, undercoupled resonators with
large quality factors can be used to store photons in the cavity on
a long time scale, with potential use as a quantum memory
\cite{Rabl2006}.

The paper is structured as follows.
In Sec.~\ref{SEC:GEOMETRY} we discuss the chosen CPW device
geometry, its fabrication and the measurement techniques used for
characterization at microwave frequencies. The dependence of the CPW
resonator frequency on the device geometry and its electrical
parameters is analyzed in Sec.~\ref{SEC:BASCIS}. In
Sec.~\ref{SEC:MODEL} the effect of the resonator coupling to an
input/output line on its quality factor, insertion loss and
resonance frequency is analyzed using a parallel LCR circuit model.
This lumped element model provides simple approximations of the
resonator properties around resonance and allows to develop an
intuitive understanding of the device. We also make use of the
transmission (or ABCD) matrix method to describe the full transmission spectrum
of the resonators and compare its predictions to our experimental
data. The characteristic properties of the higher harmonic modes of
the CPW resonators are discussed in Sec.~\ref{SEC:HARMONICS}.

\section{Device Geometry, Fabrication and Measurement Technique}
\label{SEC:GEOMETRY}

\begin{figure}[!]
\centering
\includegraphics[width = 0.95 \columnwidth]{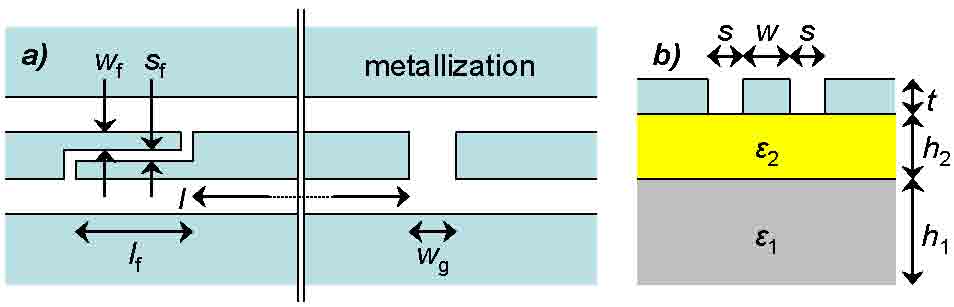}
\caption{(Color online) (a) Top view of a CPW resonator with finger capacitors
(l.h.s.) and gap capacitors (r.h.s.). (b) Cross section of a CPW
resonator design. Center conductor and lateral ground metallization
(blue) on top of a double layer substrate (grey/yellow). Parameters
are discussed in the main text.}
    \label{FIG:GEOMETRY}
\end{figure}

The planar geometry of a capacitively coupled CPW resonator is
sketched in Figure~\ref{FIG:GEOMETRY}a. The resonator is formed of a
center conductor of width $w=10\,\mu$m separated from the lateral
ground planes by a gap of width $s=6.6\,\mu$m. Resonators with
various center conductor lengths $l$ between 8 and 29\,mm aiming at
fundamental frequencies $f_0$ between 2 and 9\,GHz were designed.
These structures are easily fabricated in optical lithography while
providing sufficiently large vacuum field strengths
\cite{Schoelkopf2008}. The center conductor is coupled via gap- or
finger capacitors to the input and output transmission lines. For
small coupling capacitances gap capacitors of widths $w_{\rm{g}}=10$
to 50\,$\mu$m have been realized. To achieve larger coupling, finger
capacitors formed by from one up to eight pairs of fingers of length
$l_{\rm{f}}=100\,\mu$m, width $w_{\rm{f}}=3.3\,\mu$m and separation
$s_{\rm{f}}=3.3\,\mu$m have been designed and fabricated, see
Fig.~\ref{FIG:GEOMETRY}.

The resonators are fabricated on high resistivity, undoped, (100)-oriented,
thermally oxidized two inch silicon wafers. The oxide thickness
is $h_2=550$\,nm\,$\pm50$\,nm determined by SEM
inspection. The bulk resistivity of the Si wafer is $\rho > 3000\,\Omega\,$cm
determined at room temperature in a van-der-Pauw
measurement. The total thickness of the substrate is
$h_1=500$\,$\mu$m\,$\pm25$\,$\mu$m.  A cross-sectional sketch of the
CPW resonator is shown in Fig.~\ref{FIG:GEOMETRY}b.

The resonators were patterned in optical lithography using a one
micron thick layer of the negative tone resist ma-N 1410. The substrate
was subsequently metallized with a $t=200$\,nm\,$\pm5$\,nm thick
layer of Al, electron beam evaporated at a rate of 5\,\AA/sec and
lifted-off in $50^{\circ}$\,C hot acetone. Finally, all structures
were diced into 2\,mm$\,\times\,$7\,mm  chips, each containing an
individual resonator. The feature sizes of the fabricated devices
deviate less than $100\,$nm from the designed dimensions as
determined by SEM inspection indicating a good control over the
fabrication process.

\begin{figure}[!]
\centering
\includegraphics[width = 0.95 \columnwidth]{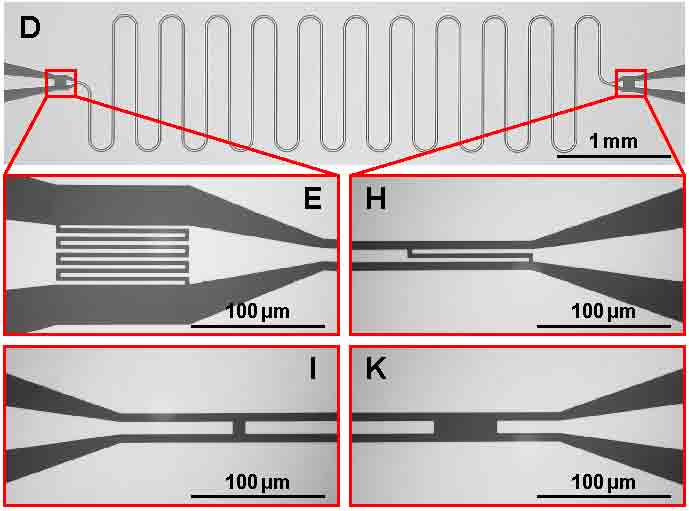}
    \caption{(Color online) Optical microscope images of an Al coplanar waveguide
    resonator (white is metallization, grey is substrate).
    The red squares in the upper image indicate the
    positions of the input/output capacitors. Also
    shown are microscope images of finger- and gap capacitor structures.
    The labels D, E, H, I, K refer to the device ID
    listed in Tab.~\ref{TAB:RESONATOR_PROPERTIES}.}
    \label{FIG:PIC_RESONATOR}
\end{figure}

Altogether, more than $80$ Al CPW resonators covering a wide range
of different coupling strengths were designed and fabricated. More
than $30$ of these devices were carefully characterized at microwave
frequencies. Figure~\ref{FIG:PIC_RESONATOR} shows optical microscope
images of the final Al resonators with different finger and gap
capacitors.

Using a 40\,GHz vector network analyzer, $S_{21}$ transmission
measurements of all resonators were performed in a pulse-tube based
dilution refrigerator system \cite{VeriCold} at temperatures of
20\,mK. The measured transmission spectra are plotted in logarithmic
units (dB) as $20 \log_{10}{|S_{21}|}$. High $Q$ resonators were
measured using a 32\,dB gain high electron mobility transistor (HEMT)
amplifier with noise temperature of $\sim$\,$5$\,K installed at
the 4\,K stage of the refrigerator as well as one or two room
temperature amplifiers with 35\,dB gain each. Low $Q$ resonators
were characterized without additional amplifiers.

The measured $Q$ of undercoupled devices can vary strongly with the
power applied to the resonator. In our measurements of high $Q$
devices the resonator transmission spectrum looses its Lorentzian
shape at drive powers above approximately $-70\,\rm{dBm}$ at the
input port of the resonator due to non-linear effects \cite{Abdo2006}.
At low drive powers, when dielectric resonator losses significantly
depend on the photon number inside the cavity \cite{Martinis2005,
Aaron2008}, measured quality factors may be substantially reduced. We
acquired $S_{21}$ transmission spectra at power levels chosen to
result in the highest measurable quality factors, i.e.~at high
enough powers to minimize dielectric loss but low enough to avoid
non-linearities. This approach has been chosen to be able to focus on
geometric properties of the resonators.

\section{Basic Resonator Properties}
\label{SEC:BASCIS}

\begin{figure}[!]
\centering
\includegraphics[]{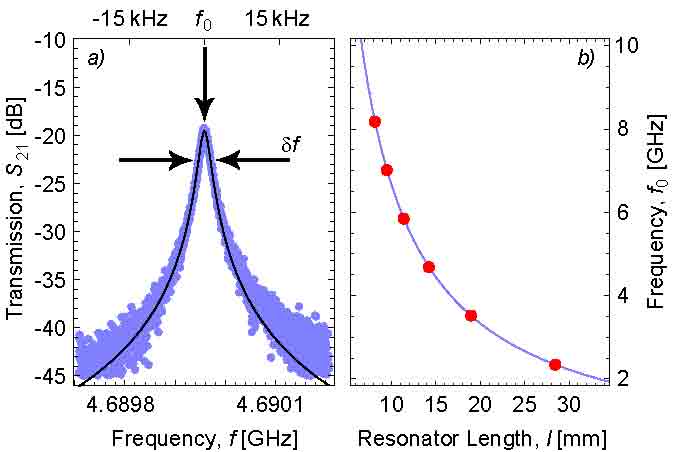}
    \caption{(Color online) (a) Transmission spectrum of a 4.7 GHz resonator.
     Data points (blue) were fitted (black) with a Lorentzian line.
     (b) Measured $f_0$ (red points) of several resonators coupled
     via $w_g=10$\,$\mu$m gap capacitors with different $l$ together
     with a fit (blue line) to the data using Eq.~(\ref{EQN:FREQUENCY})
     as fit function and $\epsilon_{\rm{eff}}$ as fit parameter.}
    \label{FIG:FREQUENCY_LENGTH}
\end{figure}

A typical transmission spectrum of a weakly gap capacitor coupled
($w_g=10\,\rm{\mu m}$) CPW resonator of length $l=14.22$\,mm is shown in
Figure~\ref{FIG:FREQUENCY_LENGTH}a. The spectrum clearly displays a
Lorentzian lineshape of width $\delta f$ centered at the resonance
frequency $f_0$. Figure~\ref{FIG:FREQUENCY_LENGTH}b shows measured
resonance frequencies $f_0$ for resonators of different length $l$,
all coupled via gap capacitors of widths $w_g=10$\,$\mu$m.
Table~\ref{TAB:RESONATOR_LENGTHS}
lists the respective values for $l$ and $f_0$. For these small
capacitors the frequency shift induced by coupling can be neglected, as
discussed in a later section. In this case the resonator`s
fundamental frequency $f_0$ is given by
\begin{equation}
f_0=\frac{c}{\sqrt{\epsilon_{\rm{eff}}}}\frac{1}{2l}.
\label{EQN:FREQUENCY}
\end{equation}
Here, ${c}/{\sqrt{\epsilon_{\rm{eff}}}}=v_{\rm{ph}}$ is the phase
velocity depending on the velocity of light in vacuum $c$ and the
effective permittivity $\epsilon_{\rm{eff}}$ of the CPW line.
$\epsilon_{\rm{eff}}$ is a function of the waveguide geometry and
the relative permittivities $\epsilon_1$ and $\epsilon_2$ of
substrate and the oxide layer, see Fig.~\ref{FIG:GEOMETRY}b.
Furthermore, $2l=\lambda_0$ is the wavelength of the fundamental
resonator mode. The length dependence of the measured resonance
frequencies $f_0$ of our samples is well described by
Eq.~(\ref{EQN:FREQUENCY}) with the effective dielectric constant
$\epsilon_{\rm{eff}}=5.05$, see Fig.~\ref{FIG:FREQUENCY_LENGTH}b.

\begin{table}
\begin{tabular}[t]{p{0.25 \columnwidth}p{0.25 \columnwidth}}
    \hline
    $f_0$ (GHz) & $l$ (mm)\\
    \hline
    2.3430 & 28.449 \\
    3.5199 & 18.970 \\
    4.6846 & 14.220 \\
    5.8491 & 11.380 \\
    7.0162 & 9.4800 \\
    8.1778 & 8.1300 \\
    \hline
\end{tabular}\caption{Designed values for resonator lengths $l$ and measured resonance frequencies $f_0$, corresponding to the data shown in Fig.~\ref{FIG:FREQUENCY_LENGTH}.}
\label{TAB:RESONATOR_LENGTHS}
\end{table}

The phase velocity
$v_{\rm{ph}}={1}/{\sqrt{L_{\rm{\ell}}C_{\rm{\ell}}}}$ of
electromagnetic waves propagating along a transmission line depends
on the capacitance $C_{\rm{\ell}}$ and inductance $L_{\rm{\ell}}$
per unit length of the line. Using conformal mapping techniques the
geometric contribution to $L_{\rm{\ell}}$ and $C_{\rm{\ell}}$ of a
CPW line is found to be \cite{Gevorgian1995, Watanabe1994}
\begin{eqnarray}
L_{\rm{\ell}} &=& \frac{\mu_0}{4}\frac{K(k_0')}{K(k_0)},
\label{EQN:IND_PER_UNIT_LENGTH} \\
C_{\rm{\ell}} &=& 4\epsilon_0\epsilon_{\rm{eff}}\frac{K(k_0)}{K(k_0')}.
\label{EQN:CAP_PER_UNIT_LENGTH}
\end{eqnarray}
\begin{table}
\begin{tabular}[t]{p{0.1 \columnwidth}p{0.25 \columnwidth}p{0.15 \columnwidth}p{0.2 \columnwidth}p{0.15 \columnwidth}}
    \hline
    ID & Coupling & $C_{\rm{\kappa}}$\,(fF) & $f_0$\,(GHz) & $Q_{\rm{L}}$\\
    \hline
    A & 8\,+\,8 finger & 56.4 & 2.2678 & $3.7\cdot10^2$ \\
    B & 7\,+\,7 finger & 48.6 & 2.2763 & $4.9\cdot10^2$ \\
    C & 6\,+\,6 finger & 42.9 & 2.2848 & $7.5\cdot10^2$ \\
    D & 5\,+\,5 finger & 35.4 & 2.2943 & $1.1\cdot10^3$ \\
    E & 4\,+\,4 finger & 26.4 & 2.3086 & $1.7\cdot10^3$ \\
    F & 3\,+\,3 finger & 18.0 & 2.3164 & $3.9\cdot10^3$ \\
    G & 2\,+\,2 finger & 11.3 & 2.3259 & $9.8\cdot10^3$ \\
    H & 1\,+\,1 finger & 3.98 & 2.3343 & $7.5\cdot10^4$ \\
    I & 10\,$\mu$m gap & 0.44 & 2.3430 & $2.0\cdot10^5$ \\
    J & 20\,$\mu$m gap & 0.38 & 2.3448 & $2.0\cdot10^5$ \\
    K & 30\,$\mu$m gap & 0.32 & 2.3459 & $2.3\cdot10^5$ \\
    L & 50\,$\mu$m gap & 0.24 & 2.3464 & $2.3\cdot10^5$ \\
    \hline
\end{tabular}
\caption{Properties of the different CPW resonators whose transmission
spectra are shown in Fig.~\ref{FIG:TRANSMISSION}. $C_{\kappa}$
denotes the simulated coupling capacitances, $f_0$ is the
measured resonance frequency and $Q_{\rm{L}}$ is the measured
quality factor.} \label{TAB:RESONATOR_PROPERTIES}
\end{table}
Here, $K$ denotes the complete elliptic integral of the first kind with the arguments
\begin{eqnarray}
k_0 &=& \frac{w}{w+2s}, \\
k_0' &=& \sqrt{1-k^2_0}.
\end{eqnarray}
For non magnetic substrates ($\mu_{\rm{eff}}=1$) and neglecting
kinetic inductance for the moment $L_{\rm{\ell}}$ is determined by
the CPW geometry only. $C_{\rm{\ell}}$ depends on the geometry and
$\epsilon_{\rm{eff}}$. Although analytical expressions for
$\epsilon_{\rm{eff}}$ exist for double layer substrates deduced from
conformal mapping \cite{Gevorgian1995}, the accuracy of these
calculations depends sensitively on the ratio between substrate
layer thicknesses and the dimensions of the CPW
cross-section\cite{Chen1997} and does not lead to accurate
predictions for our parameters. Therefore, we have calculated
$C_{\rm{\ell}}\approx 1.27\cdot10^{-10}$\,Fm$^{-1}$ using a finite
element electromagnetic simulation and values $\epsilon_1=11.6$
(see Ref.~\onlinecite{Musil1986}) for silicon and $\epsilon_2=3.78$
(see Ref.~\onlinecite{Musil1986})
for silicon oxide for our CPW geometry and substrate. From this
calculation we find $\epsilon_{\rm{eff}}\approx 5.22$ which deviates
only by about 3\% from the value extracted from our measurements.
The characteristic impedance of a CPW is then given by
$Z_0=\sqrt{L_{\rm{\ell}}/C_{\rm{\ell}}}$ which results in a value of
$59.7\,\Omega$ for our geometry. This value deviates from the
usually chosen value of $50\,\Omega$ as the original design was
optimized for a different substrate material.
\begin{figure*}[!]
\centering
\includegraphics[]{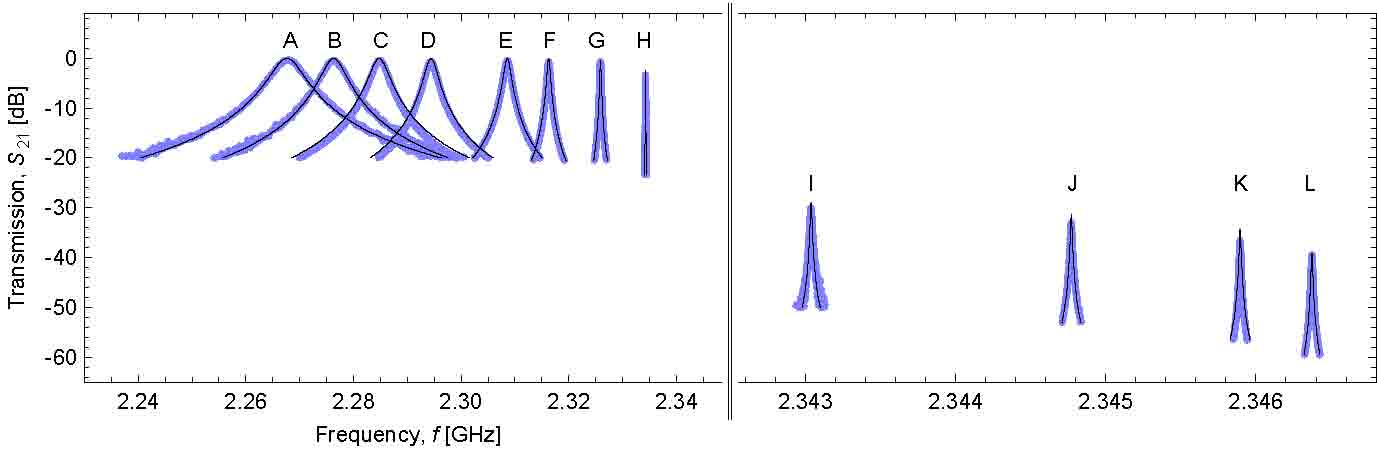}
\caption{(Color online) $S_{21}$ transmission spectra of 2.3\,GHz resonators,
symmetrically coupled to input/output lines. The left part of the
split plot shows spectra of finger capacitor coupled resonators
whereas on the right hand side one can see spectra of gap capacitor
coupled resonators. The data points (blue) were fitted (black) with the
transmission matrix method, see text.}
    \label{FIG:TRANSMISSION}
\end{figure*}

In general, for superconductors the inductance $L_{\rm{\ell}}$ is
the sum of the temperature independent geometric (magnetic)
inductance $L^{\rm{m}}_{\rm{\ell}}$ and the temperature dependent
kinetic inductance $L^{\rm{k}}_{\rm{\ell}}$ (see Ref.~\onlinecite{Tinkham1996}). For
superconductors, $L^{\rm{k}}_{\rm{\ell}}$ refers to the inertia of
moving Cooper pairs and can contribute significantly to
$L_{\rm{\ell}}$ since resistivity is suppressed and thus charge
carrier relaxation times are large. According to
Ref.~\onlinecite{Watanabe1994}, $L^{\rm{k}}_{\rm{\ell}}$ scales with
$\lambda^2(T)$, where $\lambda(T)$ is the
temperature dependent London penetration depth which can be
approximated as \cite{Watanabe1994}
$\lambda(0)=1.05\cdot10^{-3}\sqrt{\rho(T_{\rm{c}})/T_{\rm{c}}}\,\sqrt{\rm{K\,m}/\Omega}$
at zero temperature in the local and dirty limits. In the dirty (local) limit the mean free path of electrons $l_{\rm{mf}}$ is much less
than the coherence length $\xi_0=\hbar v_{\rm{f}}/\pi\Delta(0)$, where
$v_{\rm{f}}$ is the Fermi velocity of the electrons and $\Delta(0)$ is the superconducting
gap energy at zero temperature \cite{Parks1969}. The clean (nonlocal) limit occurs when $l_{\rm{mf}}$ is much larger than
$\xi_0$ (see Ref.~\onlinecite{Parks1969}). $T_{\rm{c}}=1.23\,\rm{K}$ is the critical temperature of our thin film aluminum and
$\rho(T_{\rm{c}}) = 2.06\cdot10^{-9} \, \rm{\Omega\, m}$ is the normal
state resistivity at $T=T_{\rm{c}}$. $T_{\rm{c}}$ and $\rho(T)$ were
determined in a four-point measurement of the resistance of a
lithographically patterned Al thin film meander structure from the
same substrate in dependence on temperature. The resulting residual
resistance ratio ($RRR_{\rm{300\,K/1.3\,K}}$) is $8.6$. Since our
measurements were performed at temperatures well below $T_{\rm{c}}$, $\lambda=\lambda(0)$
approximately holds and we find $\lambda(0)\approx 43\,$nm for our
Al thin films (compared to a value of $40\,$nm, given in
Ref.~\onlinecite{Poole1995}). Using the above approximation shows that
$L^{\rm{k}}_{\rm{\ell}}$ is about two orders of magnitude smaller than
$L^{\rm{m}}_{\rm{\ell}} =4.53\cdot10^{-7}$\,Hm$^{-1}$ legitimating the assumption
$L_{\rm{\ell}}\approx L^{\rm{m}}_{\rm{\ell}}$ made in
Eq.~(\ref{EQN:IND_PER_UNIT_LENGTH}). Kinetic inductance effects in
Niobium resonators are also analyzed in
Ref.~\onlinecite{Frunzio2005}.

\section{Input/Output Coupling}
\label{SEC:MODEL}

To study the effect of the capacitive coupling strength on the
microwave properties of CPW resonators, twelve 2.3\,GHz devices,
symmetrically coupled to input/output lines with different gap and
finger capacitors have been characterized, see
Table~\ref{TAB:RESONATOR_PROPERTIES} for a list of devices. The
measured transmission spectra are shown in
Fig.~\ref{FIG:TRANSMISSION}.
The left hand part of Fig.~\ref{FIG:TRANSMISSION} depicts spectra of
resonators coupled via finger capacitors having 8 down to one pairs
of fingers (devices A to H). The right hand part of
Fig.~\ref{FIG:TRANSMISSION} shows those resonators coupled via gap
capacitors with gap widths of $w_g =10$, 20, 30 and $50\,\mu$m
(devices I to L) respectively. The coupling capacitance continuously
decreases from device A to device L. The nominal values for the
coupling capacitance $C_{\rm{\kappa}}$ obtained from EM-simulations
for the investigated substrate properties and geometry are listed in
table \ref{TAB:RESONATOR_PROPERTIES}.
The resonance frequency $f_0$ and the measured quality factor
$Q_{\rm{L}}=f_0/\delta f$ of the respective device is obtained by
fitting a Lorentzian line shape
\begin{equation}
F_{\rm{Lor}}(f)=A_0\frac{\delta f}{(f-f_0)^2+\delta f^2/4},
\end{equation}
to the data, see Fig.~\ref{FIG:FREQUENCY_LENGTH}a, where $\delta f$
is the full width half maximum of the resonance. With increasing
coupling capacitance $C_{\kappa}$, Fig.~\ref{FIG:TRANSMISSION} shows
a decrease in the measured (loaded) quality factor $Q_{\rm{L}}$ and
an increase in the peak transmission, as well as a shift of $f_0$ to
lower frequencies. In the following, we demonstrate how these
characteristic resonator properties can be fully understood and
modeled consistently for the full set of data.

\begin{figure*}[!]
\centering
\includegraphics[width = 2 \columnwidth]{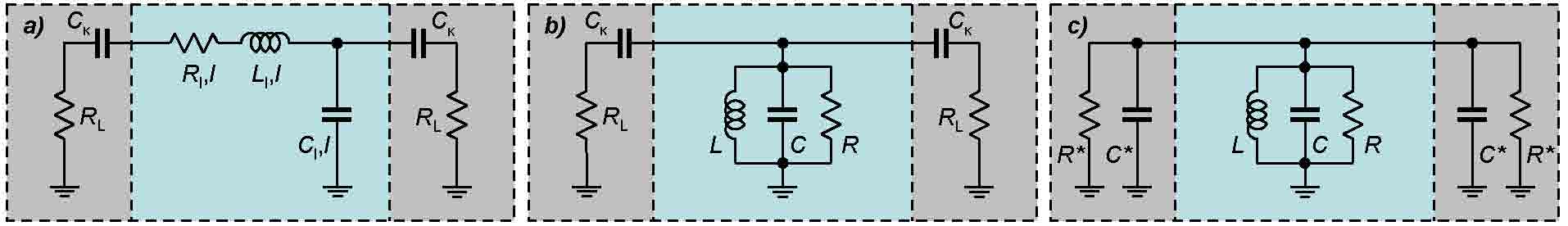}
    \caption{(Color online) (a) Distributed element representation of symmetrically coupled TL resonator.
                     (b) Parallel LCR oscillator representation of TL resonator.
                     (c) Norton equivalent of symmetrically coupled, parallel LCR oscillator. Symbols are explained in text.}
    \label{FIG:TL_REPRESENTATION}
\end{figure*}
A transmission line resonator is a distributed device with voltages
and currents varying in magnitude and phase over its length. The
distributed element representation of a symmetrically coupled
resonator is shown in Fig.~\ref{FIG:TL_REPRESENTATION}a.
$R_{\rm{\ell}}$, $L_{\rm{\ell}}$ and $C_{\rm{\ell}}$ denote the
resistance, inductance and capacitance per unit length,
respectively. According to Ref.~\onlinecite{Pozar1993} the impedance
of a TL resonator is given by
\begin{eqnarray}
Z_{\rm{TL}} & = & Z_0\frac{1+i\tan\beta l\tanh\alpha l}{\tanh\alpha l+i\tan\beta l} \\
& \approx & \frac{Z_0}{\alpha l+i\frac{\pi}{\omega_0}(\omega-\omega_{\rm{n}})}.
\label{EQN:IMPEDANCE_TL}
\end{eqnarray}
$\alpha$ is the attenuation constant and
$\beta=\omega_{\rm{n}}/v_{\rm{ph}}$ is the phase
constant of the TL. The approximation in
Eq.~(\ref{EQN:IMPEDANCE_TL}) holds when assuming small losses
($\alpha l\ll1$) and for $\omega$ close to $\omega_{\rm{n}}$. Here,
$\omega_{\rm{n}}=n\omega_0=1/\sqrt{L_{\rm{n}}C}$ is the angular
frequency of the $n$-th mode, where $n$ denotes the resonance mode
number ($n=1$ for the fundamental mode).

Around resonance, the properties of a TL resonator can be
approximated by those of a lumped element, parallel LCR oscillator,
as shown in Fig.~\ref{FIG:TL_REPRESENTATION}b, with impedance
\begin{eqnarray}
Z_{\rm{LCR}} & = & \left(\frac{1}{i\omega L_{\rm{n}}}+i\omega C + \frac{1}{R}\right)^{-1} \\
& \approx & \frac{R}{1+2iRC(\omega-\omega_{\rm{n}})},
\label{EQN:IMPEDANCE_LCR}
\end{eqnarray}
and characteristic parameters
\begin{eqnarray}
L_{\rm{n}}&=& \frac{2L_{\rm{\ell}}l}{n^2\pi^2},
\label{EQN:L_MAPPING} \\
C &=& \frac{C_{\rm{\ell}}l}{2},
\label{EQN:C_MAPPING} \\
R &=& \frac{Z_0}{\alpha l}.
\label{EQN:R_MAPPING}
\end{eqnarray}
The approximation Eq.~(\ref{EQN:IMPEDANCE_LCR}) is valid for
$\omega\approx\omega_{\rm{n}}$. The LCR model is useful to get an
intuitive understanding of the resonator properties. It simplifies
analyzing the effect of coupling the resonator to an input/output
line on the quality factor and on the resonance frequency as
discussed in the following.

\begin{figure}[!]
\centering
\includegraphics[]{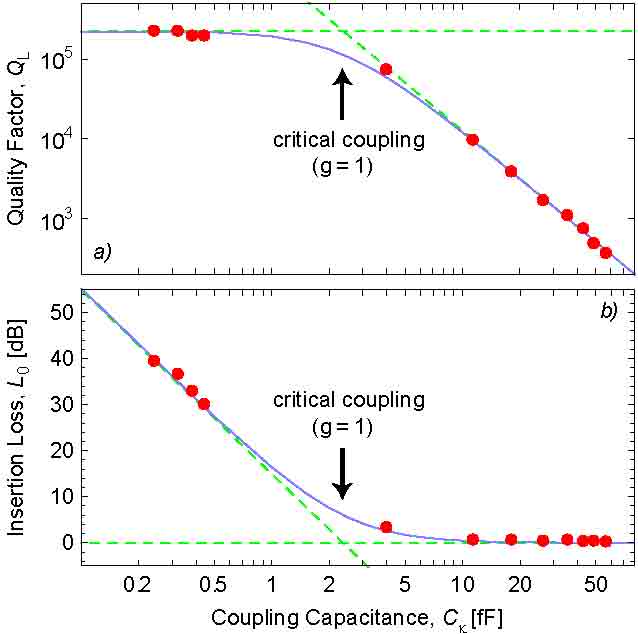}
\caption{(Color online) (a) Dependence of $Q_{\rm{L}}$ on $C_{\rm{\kappa}}$. Data
points (red) are measured quality factors. These values are compared
to $Q_{\rm{L}}$ predictions by the mapped LCR model (solid blue
line) given by Eqs.~(\ref{EQN:R_NORTON}, \ref{EQN:Q_SUM},
\ref{EQN:Q_EXT}). (b) Dependence of $L_0$ on $C_{\rm{\kappa}}$. Data
points (red) show measured $L_0$ values. The values are compared to
the mapped LCR model (solid blue line) given by
Eqs.~(\ref{EQN:R_NORTON}, \ref{EQN:Q_EXT},
\ref{EQN:INSERTION_LOSS}). Dashed lines indicate the limiting cases
for small and large coupling capacitances (see text).}
\label{FIG:QUALITY_L}
\end{figure}

The (internal) quality factor of the parallel LCR oscillator is
defined as $Q_{\rm{int}}=R\sqrt{C/L_{\rm{n}}}=\omega_{\rm{n}}RC$.
The quality factor $Q_L$ of the resonator coupled with capacitance
$C_{\kappa}$ to the input and output lines with impedance $Z_0$ is
reduced due to the resistive loading. Additionally, the frequency is
shifted because of the capacitive loading of the resonator due to
the input/output lines. To understand this effect the series
connection of $C_{\rm{\kappa}}$ and $R_{\rm{L}}$ can be transformed
into a Norton equivalent parallel connection of a resistor $R^*$ and
a capacitor $C^*$, see Figs.~\ref{FIG:TL_REPRESENTATION}b,~c, with
\begin{equation}
R^* = \frac{1+\omega^2_{\rm{n}}C^2_{\rm{\kappa}}R^2_{\rm{L}}}{\omega^2_{\rm{n}}C^2_{\rm{\kappa}}R_{\rm{L}}},
\label{EQN:R_NORTON}
\end{equation}
\begin{equation}
C^* = \frac{C_{\rm{\kappa}}}{1+\omega^2_{\rm{n}}C^2_{\rm{\kappa}}R^2_{\rm{L}}}.
\label{EQN:C_NORTON}
\end{equation}
The small capacitor $C_{\kappa}$ transforms the $R_{\rm{L}}=50\,\Omega$ load into the large impedance
$R^*=R_{\rm{L}}/k^2$ with $k=\omega_{\rm{n}}C_{\rm{\kappa}}R_{\rm{L}}\ll 1$.
For symmetric input/output coupling the loaded quality factor for
the parallel combination of $R$ and $R^*/2$ is
\begin{eqnarray}
Q_{\rm{L}} & = & \omega^*_{\rm{n}}\frac{C+2C^*}{1/R+2/R^*} \\
& \approx & \omega_{\rm{n}}\frac{C}{1/R+2/R^*}
\end{eqnarray}
with the $n$-th resonance
frequency shifted by the capacitive loading due to the parallel
combination of $C$ and $2 C^*$
\begin{equation}
\omega^*_{\rm{n}}=\frac{1}{\sqrt{L_{\rm{n}}(C+2C^*)}}.
\label{EQN:SHIFTED_FREQUENCY}
\end{equation}
For $\omega^*_{\rm{n}}\approx\omega_{\rm{n}}$ with $C+2C^*\approx
C$, the Norton equivalent expression for the
loaded quality factor $Q_{\rm{L}}$ is a parallel combination of the internal and external quality
factors
\begin{equation}
\frac{1}{Q_{\rm{L}}}=\frac{1}{Q_{\rm{int}}}+\frac{1}{Q_{\rm{ext}}},
\label{EQN:Q_SUM}
\end{equation}
with
\begin{eqnarray}
Q_{\rm{int}} &=& \omega_{\rm{n}}RC=\frac{n\pi} {2\alpha l},
\label{EQN:Q_INT} \\
Q_{\rm{ext}} &=& \frac{\omega_{\rm{n}}R^*C}{2}.
\label{EQN:Q_EXT}
\end{eqnarray}
The measured loaded quality factor $Q_{\rm{L}}$ for devices A to L
is plotted vs.~the coupling capacitance in
Fig.~\ref{FIG:QUALITY_L}a. $Q_L$ is observed to be constant for
small coupling capacitances and decreases for large ones. In the
overcoupled regime ($Q_{\rm{ext}} \ll Q_{\rm{int}}$), $Q_{\rm{L}}$
is governed by $Q_{\rm{ext}}$ which is well approximated by
$C/2\omega_{\rm{n}}R_{\rm{L}}C^2_{\rm{\kappa}}$, see dashed line in
Fig.~\ref{FIG:QUALITY_L}. Thus, in the overcoupled regime the loaded
quality factor $Q_{\rm{L}}\propto C^{-2}_{\rm{\kappa}}$ can be
controlled by the choice of the coupling capacitance. In the
undercoupled limit ($Q_{\rm{ext}} \gg Q_{\rm{int}}$) however,
$Q_{\rm{L}}$ saturates at the internal quality factor $Q_{\rm{int}}
\approx 2.3\cdot10^5$ determined by the intrinsic losses of the
resonator, see horizontal dashed line in Fig.~\ref{FIG:QUALITY_L}a.

Radiation losses are expected to be small in CPW resonators
\cite{Browne1997}, resistive losses are negligible well below the
critical temperature $T_{\rm{c}}$ of the superconductor
\cite{Frunzio2005} and at frequencies well below the superconducting
gap. We believe that dielectric losses limit the internal quality
factor of our devices, as discussed in References
\onlinecite{Martinis2005} and \onlinecite{Aaron2008}.

\begin{figure}[!]
\centering
\includegraphics[]{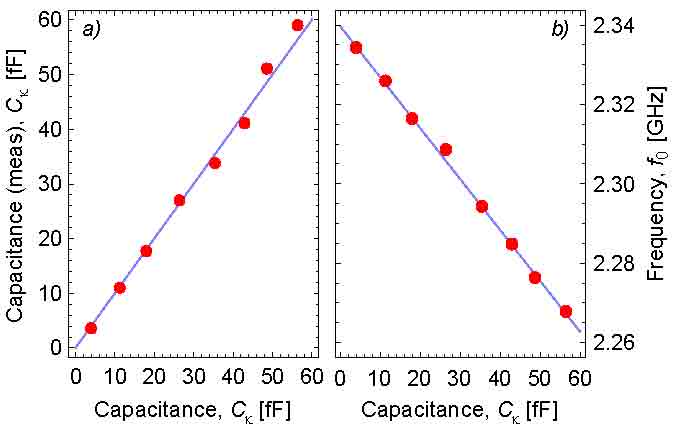}
\caption{(Color online) (a) Comparison of $C_{\rm{\kappa}}$ values extracted from
the measured quality factors using Eqs.~(\ref{EQN:R_NORTON},
\ref{EQN:Q_SUM}, \ref{EQN:Q_EXT}) to the EM-simulated values for
$C_{\rm{\kappa}}$ (red points) for devices A to H. The blue curve is
a line through origin with slope one. (b) Dependence of $f_0$ on
$C_{\rm{\kappa}}$. The mapped LCR model prediction given by
Eqs.~(\ref{EQN:C_NORTON}, \ref{EQN:SHIFTED_FREQUENCY}) is shown
(blue line) for resonators coupled via finger capacitors together
with the measured values for $f_0$ (red points).}
    \label{FIG:CAPACITANCES}
\end{figure}

Using Eqs.~(\ref{EQN:R_NORTON}, \ref{EQN:Q_SUM}, \ref{EQN:Q_EXT}),
$C_{\rm{\kappa}}$ has been extracted from the measured value of
$Q_{\rm{int}}\sim2.3\cdot10^5$ and the measured loaded quality
factors $Q_{\rm{L}}$ of the overcoupled devices A to H, see
Fig.~\ref{FIG:CAPACITANCES}. The experimental values of
$C_{\rm{\kappa}}$ are in good agreement with the ones
found from finite element calculations, listed in table
\ref{TAB:RESONATOR_PROPERTIES}, with a standard deviation of about
4\%.

The insertion loss
\begin{equation} L_0=-20\log\left(\frac{g}{g+1}\right)\,\text{dB}
\label{EQN:INSERTION_LOSS}
\end{equation}
of a resonator, i.e.~the deviation of peak transmission from unity,
is dependent on the ratio of the internal to the external quality
factor which is also called the coupling coefficient
$g=Q_{\rm{int}}/Q_{\rm{ext}}$ (see Ref.~\onlinecite{Pozar1993}). The measured values of
$L_0$ as extracted from Fig.~\ref{FIG:TRANSMISSION} are shown in
Fig.~\ref{FIG:QUALITY_L}b. For $g>1$ (large $C_{\rm{\kappa}}$) the
resonator is overcoupled and shows near unit transmission ($L_0=0$).
The resonator is said to be critically coupled for $g=1$. For $g<1$
(small $C_{\rm{\kappa}}$) the resonator is undercoupled and the
transmission is significantly reduced. In this case $L_0$ is well
approximated by
$-20\log(2\omega_{\rm{n}}Q_{\rm{int}}R_{\rm{L}}C^2_{\rm{\kappa}}/C)$,
see dashed line in Fig.~\ref{FIG:QUALITY_L}b, as calculated from
Eqs.~(\ref{EQN:R_NORTON}, \ref{EQN:Q_EXT},
\ref{EQN:INSERTION_LOSS}). $Q_{\rm{ext}}$ and $Q_{\rm{int}}$ can be
determined from $Q_{\rm{L}}$ and $L_0$ using Eqs.~(\ref{EQN:Q_SUM},
\ref{EQN:INSERTION_LOSS}), thus allowing to roughly estimate
internal losses even of an overcoupled cavity.\\
For the overcoupled devices A to H the coupling induced resonator
frequency shift as extracted from Fig.~\ref{FIG:TRANSMISSION} is in
good agreement with calculations based on
Eqs.~(\ref{EQN:C_NORTON}, \ref{EQN:SHIFTED_FREQUENCY}), see Fig.~\ref{FIG:CAPACITANCES}b.
For $C^*\approx C_{\kappa}$ and $C\gg C_{\kappa}$ one can
Taylor-approximate $\omega^*_{\rm{n}}$ as
$\omega_{\rm{n}}(1-C_{\kappa}/C)$. As a result the relative
resonator frequency shift is
$(\omega^*_{\rm{n}}-\omega_{\rm{n}})/\omega_{\rm{n}}=-C_{\kappa}/C$
for symmetric coupling. Figure \ref{FIG:CAPACITANCES}b shows the
expected linear dependence with a maximum frequency shift of about 3\% over a
range of 60\,fF in $C_{\rm{\kappa}}$.\\

As an alternative method to the LCR model which is only an accurate description near resonance
we have analyzed our data using the transmission matrix method \cite{Pozar1993}.
Using this method the full transmission spectrum of the CPW resonator can be calculated.
However, because of the mathematical structure of the model it is more involved to gain intuitive
understanding of the CPW devices.

All measured $S_{21}$ transmission spectra are consistently fit
with a single set of parameters, see Fig.~\ref{FIG:TRANSMISSION}.
The transmission or ABCD matrix of a symmetrically coupled TL is defined by the
product of an input-, a transmission-, and an output matrix as
\begin{equation}
\left(
\begin{matrix}
    A & B\\
    C & D
\end{matrix}
\right)= \left(
\begin{matrix}
    1 & Z_{\rm{in}}\\
    0 & 1
\end{matrix}
\right) \left(
\begin{matrix}
    t_{11} & t_{12}\\
    t_{21} & t_{22}
\end{matrix}
\right) \left(
\begin{matrix}
    1 & Z_{\rm{out}}\\
    0 & 1
\end{matrix}
\right), \label{EQN:ABCD_MATRIX}
\end{equation}
with input/output impedances $Z_{\rm{in/out}}=1/i\omega
C_{\rm{\kappa}}$ and the transmission matrix parameters
\begin{eqnarray}
t_{11}&=&\cosh{(\gamma l)},\\
t_{12}&=&Z_0\sinh{(\gamma l)},\\
t_{21}&=&1/Z_0\sinh{(\gamma l)},\\
t_{22}&=&\cosh{(\gamma l)}.
\end{eqnarray}
Here, $\gamma=\alpha+i\beta$ is the TL wave propagation coefficient.
The resonator transmission spectrum is then defined by the ABCD
matrix components as
\begin{equation}
S_{21}=\frac{2}{A+B/R_{\rm{L}}+CR_{\rm{L}}+D}.
\label{EQN:S21_SPECTRUM}
\end{equation}
Here, $R_{\rm{L}}$ is the real part of the load impedance,
accounting for outer circuit components.
$\alpha$ is determined by $Q_{\rm{int}}$ and $l$ and $\beta$ depends
on $\epsilon_{\rm{eff}}$ as discussed before. According to
Eqs.~(\ref{EQN:IND_PER_UNIT_LENGTH}, \ref{EQN:CAP_PER_UNIT_LENGTH})
$Z_0$ is determined by $\epsilon_{\rm{eff}}$, $w$ and $s$. The
attenuation constant is $\alpha \sim 2.4\cdot10^{-4}\,\rm{m}^{-1}$ as determined
from $Q_{\rm{int}} \sim 2.3\cdot10^5$.

For gap capacitor coupled devices, the measured data fits very well,
see Fig.~\ref{FIG:TRANSMISSION}, to the transmission spectrum
calculated using the ABCD matrix method with
$\overline{\epsilon}_{\rm{eff}}=5.05$, already obtained from the
measured dependence of $f_0$ on the resonator length, see
Fig.~\ref{FIG:FREQUENCY_LENGTH}. For finger capacitor coupled
structures however, see Fig.~\ref{FIG:GEOMETRY}a, approximately 40\%
of the length of each 100\,$\mu$m finger has to be added to the
length $l$ of the bare resonators in order to obtain good fits to
the resonance frequency $f_0$. This result is independent of the
number of fingers. The ABCD matrix model describes the full
transmission spectra of all measured devices very well with a single
set of parameters, see Fig.~\ref{FIG:TRANSMISSION}.

\section{Harmonic Modes}
\label{SEC:HARMONICS}

\begin{figure}[!]
\centering
\includegraphics[]{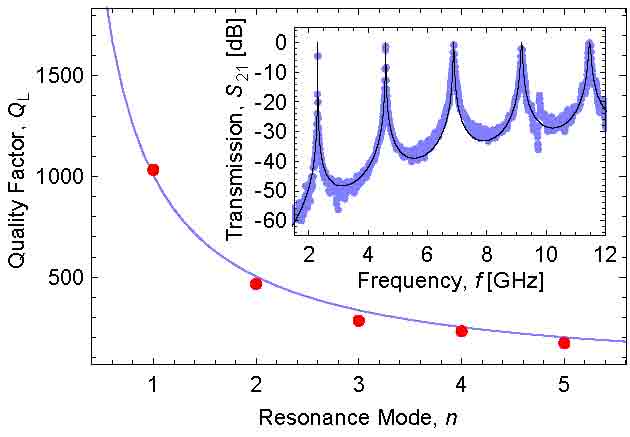}
\caption{Measured quality factors for the overcoupled resonator D
vs.~mode number $n$ (red points) together with the prediction of the
mapped LCR model given by Eqs.~(\ref{EQN:SHIFTED_FREQUENCY},
\ref{EQN:Q_EXT}) (solid blue line). The inset shows the $S_{21}$
transmission spectrum of resonator D with fundamental mode and
harmonics. The measured data (blue) is compared to the $S_{21}$
spectrum (black) obtained by the ABCD matrix method.}
    \label{FIG:QUALITY_L_HARMONICS}
\end{figure}

So far we have only discussed the properties of the fundamental
resonance frequency of any of the measured resonators. A full
transmission spectrum of the overcoupled resonator D, including 5
harmonic modes, is shown in Fig.~\ref{FIG:QUALITY_L_HARMONICS}.
The measured spectrum fits well to the ABCD matrix model for the
fundamental frequency and also for higher cavity modes, displaying a
decrease of the loaded quality factor with harmonic number. The
dependence of the measured quality factor $Q_{\rm{L}}$ on the mode
number $n$ is in good agreement with Eqs.~(\ref{EQN:Q_SUM},
\ref{EQN:Q_EXT}) and scales approximately as
$C/2n\omega_0R_{\rm{L}}C^2_{\kappa}$.

\section{Conclusions}
In summary, we have designed and fabricated symmetrically coupled
coplanar waveguide resonators over a wide range of resonance
frequencies and coupling strengths. We demonstrate that loaded
quality factors and resonance frequencies can be controlled and that
the LCR- and ABCD matrix models are in good agreement with measured
data for fundamental and harmonic modes. In the case of resonators
coupled via finger capacitors simulated values for $C_{\rm{\kappa}}$
deviate by only about 4\%. About $40\%$ of the capacitor finger length
has to be added to the total resonator length to obtain a good fit
to the resonance frequency.

The resonator properties
discussed above are consistent with those obtained from measurements
of additional devices with fundamental frequencies of 3.5, 4.7, 5.8, 7.0 and
8.2\,GHz. The experimental results presented in this paper were obtained
for Al based resonators on an oxidized silicon substrate. The
methods of analysis should also be applicable to CPW devices fabricated on different
substrates and with different superconducting materials.
The good understanding of geometric and electrical properties of CPW resonators
will certainly foster further research on their use as radiation detectors,
in quantum electrodynamics and quantum information processing applications.

\begin{acknowledgments}
We thank P.~Fallahi for designing the optical lithography mask used for fabricating the devices and
L.~Frunzio and R.~Schoelkopf at Yale University for
their continued collaboration on resonator fabrication and characterization. We also thank
the Yale group and the group of M.~Siegel at the Institute for Micro- and Nanoelectronic Systems
at University of Karlsruhe for exchange of materials and preparation of thin films. Furthermore
we acknowledge the ETH Z\"urich FIRST Center for Micro- and Nanoscience for providing and
supporting the device fabrication infrastructure essential for this project. We acknowledge discussions with
J.~Martinis and K.~Lehnert and thank D.~Schuster for valuable comments on the manuscript.
This work was supported by Swiss National Fund (SNF) and ETH Z\"urich. P.~J.~L.~was supported by the EC with a MC-EIF.
\end{acknowledgments}


\begin{thebibliography}{42}
\expandafter\ifx\csname natexlab\endcsname\relax\def\natexlab#1{#1}\fi
\expandafter\ifx\csname bibnamefont\endcsname\relax
  \def\bibnamefont#1{#1}\fi
\expandafter\ifx\csname bibfnamefont\endcsname\relax
  \def\bibfnamefont#1{#1}\fi
\expandafter\ifx\csname citenamefont\endcsname\relax
  \def\citenamefont#1{#1}\fi
\expandafter\ifx\csname url\endcsname\relax
  \def\url#1{\texttt{#1}}\fi
\expandafter\ifx\csname urlprefix\endcsname\relax\def\urlprefix{URL }\fi
\providecommand{\bibinfo}[2]{#2}
\providecommand{\eprint}[2][]{\url{#2}}

\bibitem[{\citenamefont{Mazin et~al.}(2002)\citenamefont{Mazin, Day, LeDuc,
  Vayonakis, and Zmuidzinas}}]{Mazin2002}
\bibinfo{author}{\bibfnamefont{B.~A.} \bibnamefont{Mazin}},
  \bibinfo{author}{\bibfnamefont{P.~K.} \bibnamefont{Day}},
  \bibinfo{author}{\bibfnamefont{H.~G.} \bibnamefont{LeDuc}},
  \bibinfo{author}{\bibfnamefont{A.}~\bibnamefont{Vayonakis}},
  \bibnamefont{and}
  \bibinfo{author}{\bibfnamefont{J.}~\bibnamefont{Zmuidzinas}},
  \bibinfo{journal}{2002 Proc. SPIE} \textbf{\bibinfo{volume}{4849}},
  \bibinfo{pages}{283} (\bibinfo{year}{2002}).

\bibitem[{\citenamefont{Day et~al.}(2003)\citenamefont{Day, LeDuc, Mazin,
  Vayonakis, and Zmuidzinas}}]{Day2003}
\bibinfo{author}{\bibfnamefont{P.~K.} \bibnamefont{Day}},
  \bibinfo{author}{\bibfnamefont{H.~G.} \bibnamefont{LeDuc}},
  \bibinfo{author}{\bibfnamefont{B.~A.} \bibnamefont{Mazin}},
  \bibinfo{author}{\bibfnamefont{A.}~\bibnamefont{Vayonakis}},
  \bibnamefont{and}
  \bibinfo{author}{\bibfnamefont{J.}~\bibnamefont{Zmuidzinas}},
  \bibinfo{journal}{Nature} \textbf{\bibinfo{volume}{425}},
  \bibinfo{pages}{817} (\bibinfo{year}{2003}).

\bibitem[{\citenamefont{Zmuidzinas and Richards}(2004)}]{Zmuidzinas2004}
\bibinfo{author}{\bibfnamefont{J.}~\bibnamefont{Zmuidzinas}} \bibnamefont{and}
  \bibinfo{author}{\bibfnamefont{P.~L.} \bibnamefont{Richards}},
  \bibinfo{journal}{Proc. IEEE} \textbf{\bibinfo{volume}{92(10)}},
  \bibinfo{pages}{1597} (\bibinfo{year}{2004}).

\bibitem[{\citenamefont{Mazin et~al.}(2008)\citenamefont{Mazin, Eckart, Bumble,
  Golwala, Day, Gao, and Zmuidzinas}}]{Mazin2008}
\bibinfo{author}{\bibfnamefont{B.~A.} \bibnamefont{Mazin}},
  \bibinfo{author}{\bibfnamefont{M.~E.} \bibnamefont{Eckart}},
  \bibinfo{author}{\bibfnamefont{B.}~\bibnamefont{Bumble}},
  \bibinfo{author}{\bibfnamefont{S.}~\bibnamefont{Golwala}},
  \bibinfo{author}{\bibfnamefont{P.}~\bibnamefont{Day}},
  \bibinfo{author}{\bibfnamefont{J.}~\bibnamefont{Gao}}, \bibnamefont{and}
  \bibinfo{author}{\bibfnamefont{J.}~\bibnamefont{Zmuidzinas}},
  \bibinfo{journal}{J. Low Temp. Phys.} \textbf{\bibinfo{volume}{151}},
  \bibinfo{pages}{537} (\bibinfo{year}{2008}).

\bibitem[{\citenamefont{Vardulakis et~al.}(2008)\citenamefont{Vardulakis,
  Withington, Goldie, and Glowacka}}]{Vardulakis2008}
\bibinfo{author}{\bibfnamefont{G.}~\bibnamefont{Vardulakis}},
  \bibinfo{author}{\bibfnamefont{S.}~\bibnamefont{Withington}},
  \bibinfo{author}{\bibfnamefont{D.~J.} \bibnamefont{Goldie}},
  \bibnamefont{and} \bibinfo{author}{\bibfnamefont{D.~M.}
  \bibnamefont{Glowacka}}, \bibinfo{journal}{Meas. Sci. Technol.}
  \textbf{\bibinfo{volume}{19}}, \bibinfo{pages}{015509}
  (\bibinfo{year}{2008}).

\bibitem[{\citenamefont{Thol\'{e}n et~al.}(2007)\citenamefont{Thol\'{e}n,
  Erg\"ul, Doherty, Weber, Gr\'{e}gis, and Haviland}}]{Tholen2007}
\bibinfo{author}{\bibfnamefont{E.}~\bibnamefont{Thol\'{e}n}},
  \bibinfo{author}{\bibfnamefont{A.}~\bibnamefont{Erg\"ul}},
  \bibinfo{author}{\bibfnamefont{E.}~\bibnamefont{Doherty}},
  \bibinfo{author}{\bibfnamefont{F.}~\bibnamefont{Weber}},
  \bibinfo{author}{\bibfnamefont{F.}~\bibnamefont{Gr\'{e}gis}},
  \bibnamefont{and} \bibinfo{author}{\bibfnamefont{D.}~\bibnamefont{Haviland}},
  \bibinfo{journal}{Appl. Phys. Lett.} \textbf{\bibinfo{volume}{90}},
  \bibinfo{pages}{253509} (\bibinfo{year}{2007}).

\bibitem[{\citenamefont{Castellanos-Beltran and Lehnert}(2007)}]{Lehnert2007}
\bibinfo{author}{\bibfnamefont{M.}~\bibnamefont{Castellanos-Beltran}}
  \bibnamefont{and} \bibinfo{author}{\bibfnamefont{K.}~\bibnamefont{Lehnert}},
  \bibinfo{journal}{Appl. Phys. Lett.} \textbf{\bibinfo{volume}{91}},
  \bibinfo{pages}{083509} (\bibinfo{year}{2007}).

\bibitem[{\citenamefont{Bergeal et~al.}(2008)\citenamefont{Bergeal, Vijay,
  Manucharyan, Siddiqi, Schoelkopf, Girvin, and Devoret}}]{Bergeal2008}
\bibinfo{author}{\bibfnamefont{N.}~\bibnamefont{Bergeal}},
  \bibinfo{author}{\bibfnamefont{R.}~\bibnamefont{Vijay}},
  \bibinfo{author}{\bibfnamefont{V.~E.} \bibnamefont{Manucharyan}},
  \bibinfo{author}{\bibfnamefont{I.}~\bibnamefont{Siddiqi}},
  \bibinfo{author}{\bibfnamefont{R.~J.} \bibnamefont{Schoelkopf}},
  \bibinfo{author}{\bibfnamefont{S.~M.} \bibnamefont{Girvin}},
  \bibnamefont{and} \bibinfo{author}{\bibfnamefont{M.~H.}
  \bibnamefont{Devoret}}, \bibinfo{journal}{arXiv.org:0805.3452}
  (\bibinfo{year}{2008}).

\bibitem[{\citenamefont{Palacios-Laloy
  et~al.}(2007)\citenamefont{Palacios-Laloy, Nguyen, Mallet, Bertet, Vion, and
  Esteve}}]{Palacios-Laloy2007}
\bibinfo{author}{\bibfnamefont{A.}~\bibnamefont{Palacios-Laloy}},
  \bibinfo{author}{\bibfnamefont{F.}~\bibnamefont{Nguyen}},
  \bibinfo{author}{\bibfnamefont{F.}~\bibnamefont{Mallet}},
  \bibinfo{author}{\bibfnamefont{P.}~\bibnamefont{Bertet}},
  \bibinfo{author}{\bibfnamefont{D.}~\bibnamefont{Vion}}, \bibnamefont{and}
  \bibinfo{author}{\bibfnamefont{D.}~\bibnamefont{Esteve}},
  \bibinfo{journal}{arXiv:0712.0221v1}  (\bibinfo{year}{2007}).

\bibitem[{\citenamefont{Sandberg et~al.}(2008)\citenamefont{Sandberg, Wilson,
  Persson, Johansson, Shumeiko, Duty, and Delsing}}]{Sandberg2008}
\bibinfo{author}{\bibfnamefont{M.}~\bibnamefont{Sandberg}},
  \bibinfo{author}{\bibfnamefont{C.}~\bibnamefont{Wilson}},
  \bibinfo{author}{\bibfnamefont{F.}~\bibnamefont{Persson}},
  \bibinfo{author}{\bibfnamefont{G.}~\bibnamefont{Johansson}},
  \bibinfo{author}{\bibfnamefont{V.}~\bibnamefont{Shumeiko}},
  \bibinfo{author}{\bibfnamefont{T.}~\bibnamefont{Duty}}, \bibnamefont{and}
  \bibinfo{author}{\bibfnamefont{P.}~\bibnamefont{Delsing}},
  \bibinfo{journal}{arXiv:0801.2479v1}  (\bibinfo{year}{2008}).

\bibitem[{\citenamefont{Wallraff et~al.}(2004)\citenamefont{Wallraff, Schuster,
  Blais, Frunzio, Huang, Majer, Kumar, Girvin, and Schoelkopf}}]{Wallraff2004a}
\bibinfo{author}{\bibfnamefont{A.}~\bibnamefont{Wallraff}},
  \bibinfo{author}{\bibfnamefont{D.~I.} \bibnamefont{Schuster}},
  \bibinfo{author}{\bibfnamefont{A.}~\bibnamefont{Blais}},
  \bibinfo{author}{\bibfnamefont{L.}~\bibnamefont{Frunzio}},
  \bibinfo{author}{\bibfnamefont{R.-S.} \bibnamefont{Huang}},
  \bibinfo{author}{\bibfnamefont{J.}~\bibnamefont{Majer}},
  \bibinfo{author}{\bibfnamefont{S.}~\bibnamefont{Kumar}},
  \bibinfo{author}{\bibfnamefont{S.~M.} \bibnamefont{Girvin}},
  \bibnamefont{and} \bibinfo{author}{\bibfnamefont{R.~J.}
  \bibnamefont{Schoelkopf}}, \bibinfo{journal}{Nature}
  \textbf{\bibinfo{volume}{431}}, \bibinfo{pages}{162} (\bibinfo{year}{2004}).

\bibitem[{\citenamefont{Wallraff et~al.}(2005)\citenamefont{Wallraff, Schuster,
  Blais, Frunzio, Majer, Girvin, and Schoelkopf}}]{Wallraff2005}
\bibinfo{author}{\bibfnamefont{A.}~\bibnamefont{Wallraff}},
  \bibinfo{author}{\bibfnamefont{D.~I.} \bibnamefont{Schuster}},
  \bibinfo{author}{\bibfnamefont{A.}~\bibnamefont{Blais}},
  \bibinfo{author}{\bibfnamefont{L.}~\bibnamefont{Frunzio}},
  \bibinfo{author}{\bibfnamefont{J.}~\bibnamefont{Majer}},
  \bibinfo{author}{\bibfnamefont{S.~M.} \bibnamefont{Girvin}},
  \bibnamefont{and} \bibinfo{author}{\bibfnamefont{R.~J.}
  \bibnamefont{Schoelkopf}}, \bibinfo{journal}{Phys. Rev. Lett.}
  \textbf{\bibinfo{volume}{95}}, \bibinfo{pages}{060501}
  (\bibinfo{year}{2005}).

\bibitem[{\citenamefont{Schuster et~al.}(2005)\citenamefont{Schuster, Wallraff,
  Blais, Frunzio, Huang, Majer, Girvin, and Schoelkopf}}]{Schuster2005}
\bibinfo{author}{\bibfnamefont{D.~I.} \bibnamefont{Schuster}},
  \bibinfo{author}{\bibfnamefont{A.}~\bibnamefont{Wallraff}},
  \bibinfo{author}{\bibfnamefont{A.}~\bibnamefont{Blais}},
  \bibinfo{author}{\bibfnamefont{L.}~\bibnamefont{Frunzio}},
  \bibinfo{author}{\bibfnamefont{R.-S.} \bibnamefont{Huang}},
  \bibinfo{author}{\bibfnamefont{J.}~\bibnamefont{Majer}},
  \bibinfo{author}{\bibfnamefont{S.~M.} \bibnamefont{Girvin}},
  \bibnamefont{and} \bibinfo{author}{\bibfnamefont{R.~J.}
  \bibnamefont{Schoelkopf}}, \bibinfo{journal}{Phys. Rev. Lett.}
  \textbf{\bibinfo{volume}{94}}, \bibinfo{pages}{123602}
  (\bibinfo{year}{2005}).

\bibitem[{\citenamefont{Wallraff et~al.}(2007)\citenamefont{Wallraff, Schuster,
  Blais, Gambetta, Schreier, Frunzio, Devoret, Girvin, and
  Schoelkopf}}]{Wallraff2007}
\bibinfo{author}{\bibfnamefont{A.}~\bibnamefont{Wallraff}},
  \bibinfo{author}{\bibfnamefont{D.~I.} \bibnamefont{Schuster}},
  \bibinfo{author}{\bibfnamefont{A.}~\bibnamefont{Blais}},
  \bibinfo{author}{\bibfnamefont{J.~M.} \bibnamefont{Gambetta}},
  \bibinfo{author}{\bibfnamefont{J.}~\bibnamefont{Schreier}},
  \bibinfo{author}{\bibfnamefont{L.}~\bibnamefont{Frunzio}},
  \bibinfo{author}{\bibfnamefont{M.~H.} \bibnamefont{Devoret}},
  \bibinfo{author}{\bibfnamefont{S.~M.} \bibnamefont{Girvin}},
  \bibnamefont{and} \bibinfo{author}{\bibfnamefont{R.~J.}
  \bibnamefont{Schoelkopf}}, \bibinfo{journal}{Phys. Rev. Lett.}
  \textbf{\bibinfo{volume}{99}}, \bibinfo{pages}{050501}
  (\bibinfo{year}{2007}).

\bibitem[{\citenamefont{Schuster
  et~al.}(2007{\natexlab{a}})\citenamefont{Schuster, Houck, Schreier, Wallraff,
  Gambetta, Blais, Frunzio, Majer, Johnson, Devoret et~al.}}]{Schuster2007a}
\bibinfo{author}{\bibfnamefont{D.~I.} \bibnamefont{Schuster}},
  \bibinfo{author}{\bibfnamefont{A.~A.} \bibnamefont{Houck}},
  \bibinfo{author}{\bibfnamefont{J.~A.} \bibnamefont{Schreier}},
  \bibinfo{author}{\bibfnamefont{A.}~\bibnamefont{Wallraff}},
  \bibinfo{author}{\bibfnamefont{J.~M.} \bibnamefont{Gambetta}},
  \bibinfo{author}{\bibfnamefont{A.}~\bibnamefont{Blais}},
  \bibinfo{author}{\bibfnamefont{L.}~\bibnamefont{Frunzio}},
  \bibinfo{author}{\bibfnamefont{J.}~\bibnamefont{Majer}},
  \bibinfo{author}{\bibfnamefont{B.}~\bibnamefont{Johnson}},
  \bibinfo{author}{\bibfnamefont{M.~H.} \bibnamefont{Devoret}},
  \bibnamefont{et~al.}, \bibinfo{journal}{Nature}
  \textbf{\bibinfo{volume}{445}}, \bibinfo{pages}{515}
  (\bibinfo{year}{2007}{\natexlab{a}}).

\bibitem[{\citenamefont{Schuster
  et~al.}(2007{\natexlab{b}})\citenamefont{Schuster, Wallraff, Blais, Frunzio,
  Huang, Majer, Girvin, and Schoelkopf}}]{Schuster2007b}
\bibinfo{author}{\bibfnamefont{D.~I.} \bibnamefont{Schuster}},
  \bibinfo{author}{\bibfnamefont{A.}~\bibnamefont{Wallraff}},
  \bibinfo{author}{\bibfnamefont{A.}~\bibnamefont{Blais}},
  \bibinfo{author}{\bibfnamefont{L.}~\bibnamefont{Frunzio}},
  \bibinfo{author}{\bibfnamefont{R.~S.} \bibnamefont{Huang}},
  \bibinfo{author}{\bibfnamefont{J.}~\bibnamefont{Majer}},
  \bibinfo{author}{\bibfnamefont{S.~M.} \bibnamefont{Girvin}},
  \bibnamefont{and} \bibinfo{author}{\bibfnamefont{R.~J.}
  \bibnamefont{Schoelkopf}}, \bibinfo{journal}{Phys. Rev. Lett.}
  \textbf{\bibinfo{volume}{98}}, \bibinfo{pages}{049902}
  (\bibinfo{year}{2007}{\natexlab{b}}).

\bibitem[{\citenamefont{Majer et~al.}(2007)\citenamefont{Majer, Chow, Gambetta,
  Koch, Johnson, Schreier, Frunzio, Schuster, Houck, Wallraff
  et~al.}}]{Majer2007}
\bibinfo{author}{\bibfnamefont{J.}~\bibnamefont{Majer}},
  \bibinfo{author}{\bibfnamefont{J.~M.} \bibnamefont{Chow}},
  \bibinfo{author}{\bibfnamefont{J.~M.} \bibnamefont{Gambetta}},
  \bibinfo{author}{\bibfnamefont{J.}~\bibnamefont{Koch}},
  \bibinfo{author}{\bibfnamefont{B.~R.} \bibnamefont{Johnson}},
  \bibinfo{author}{\bibfnamefont{J.~A.} \bibnamefont{Schreier}},
  \bibinfo{author}{\bibfnamefont{L.}~\bibnamefont{Frunzio}},
  \bibinfo{author}{\bibfnamefont{D.~I.} \bibnamefont{Schuster}},
  \bibinfo{author}{\bibfnamefont{A.~A.} \bibnamefont{Houck}},
  \bibinfo{author}{\bibfnamefont{A.}~\bibnamefont{Wallraff}},
  \bibnamefont{et~al.}, \bibinfo{journal}{Nature}
  \textbf{\bibinfo{volume}{449}}, \bibinfo{pages}{443} (\bibinfo{year}{2007}).

\bibitem[{\citenamefont{Houck et~al.}(2007)\citenamefont{Houck, Schuster,
  Gambetta, Schreier, Johnson, Chow, Frunzio, Majer, Devoret, Girvin
  et~al.}}]{Houck2007}
\bibinfo{author}{\bibfnamefont{A.~A.} \bibnamefont{Houck}},
  \bibinfo{author}{\bibfnamefont{D.~I.} \bibnamefont{Schuster}},
  \bibinfo{author}{\bibfnamefont{J.~M.} \bibnamefont{Gambetta}},
  \bibinfo{author}{\bibfnamefont{J.~A.} \bibnamefont{Schreier}},
  \bibinfo{author}{\bibfnamefont{B.~R.} \bibnamefont{Johnson}},
  \bibinfo{author}{\bibfnamefont{J.~M.} \bibnamefont{Chow}},
  \bibinfo{author}{\bibfnamefont{L.}~\bibnamefont{Frunzio}},
  \bibinfo{author}{\bibfnamefont{J.}~\bibnamefont{Majer}},
  \bibinfo{author}{\bibfnamefont{M.~H.} \bibnamefont{Devoret}},
  \bibinfo{author}{\bibfnamefont{S.~M.} \bibnamefont{Girvin}},
  \bibnamefont{et~al.}, \bibinfo{journal}{Nature}
  \textbf{\bibinfo{volume}{449}}, \bibinfo{pages}{328} (\bibinfo{year}{2007}).

\bibitem[{\citenamefont{Astafiev et~al.}(2007)\citenamefont{Astafiev, Inomata,
  Niskanen, Yamamoto, Pashkin, Nakamura, and Tsai}}]{Astafiev2007}
\bibinfo{author}{\bibfnamefont{O.}~\bibnamefont{Astafiev}},
  \bibinfo{author}{\bibfnamefont{K.}~\bibnamefont{Inomata}},
  \bibinfo{author}{\bibfnamefont{A.~O.} \bibnamefont{Niskanen}},
  \bibinfo{author}{\bibfnamefont{T.}~\bibnamefont{Yamamoto}},
  \bibinfo{author}{\bibfnamefont{Y.~A.} \bibnamefont{Pashkin}},
  \bibinfo{author}{\bibfnamefont{Y.}~\bibnamefont{Nakamura}}, \bibnamefont{and}
  \bibinfo{author}{\bibfnamefont{J.~S.} \bibnamefont{Tsai}},
  \bibinfo{journal}{Nature} \textbf{\bibinfo{volume}{449}},
  \bibinfo{pages}{588} (\bibinfo{year}{2007}).

\bibitem[{\citenamefont{Sillanp\"{a}\"{a}
  et~al.}(2007)\citenamefont{Sillanp\"{a}\"{a}, Park, and
  Simmonds}}]{Sillanpaa2007}
\bibinfo{author}{\bibfnamefont{M.~A.} \bibnamefont{Sillanp\"{a}\"{a}}},
  \bibinfo{author}{\bibfnamefont{J.~I.} \bibnamefont{Park}}, \bibnamefont{and}
  \bibinfo{author}{\bibfnamefont{R.~W.} \bibnamefont{Simmonds}},
  \bibinfo{journal}{Nature} \textbf{\bibinfo{volume}{449}},
  \bibinfo{pages}{438} (\bibinfo{year}{2007}).

\bibitem[{\citenamefont{Hofheinz et~al.}(454)\citenamefont{Hofheinz, Weig,
  Ansmann, Bialczak, Lucero, Neeley, O'Connell, Wang, Martinis, and
  Cleland}}]{Hofheinz2008}
\bibinfo{author}{\bibfnamefont{M.}~\bibnamefont{Hofheinz}},
  \bibinfo{author}{\bibfnamefont{E.~M.} \bibnamefont{Weig}},
  \bibinfo{author}{\bibfnamefont{M.}~\bibnamefont{Ansmann}},
  \bibinfo{author}{\bibfnamefont{R.~C.} \bibnamefont{Bialczak}},
  \bibinfo{author}{\bibfnamefont{E.}~\bibnamefont{Lucero}},
  \bibinfo{author}{\bibfnamefont{M.}~\bibnamefont{Neeley}},
  \bibinfo{author}{\bibfnamefont{A.~D.} \bibnamefont{O'Connell}},
  \bibinfo{author}{\bibfnamefont{H.}~\bibnamefont{Wang}},
  \bibinfo{author}{\bibfnamefont{J.~M.} \bibnamefont{Martinis}},
  \bibnamefont{and} \bibinfo{author}{\bibfnamefont{A.~N.}
  \bibnamefont{Cleland}}, \bibinfo{journal}{Nature} p. \bibinfo{pages}{310}
  (\bibinfo{year}{454}).

\bibitem[{\citenamefont{Fink et~al.}(2008)\citenamefont{Fink, G\"oppl, Baur,
  Bianchetti, Leek, Blais, and Wallraff}}]{Fink2008}
\bibinfo{author}{\bibfnamefont{J.~M.} \bibnamefont{Fink}},
  \bibinfo{author}{\bibfnamefont{M.}~\bibnamefont{G\"oppl}},
  \bibinfo{author}{\bibfnamefont{M.}~\bibnamefont{Baur}},
  \bibinfo{author}{\bibfnamefont{R.}~\bibnamefont{Bianchetti}},
  \bibinfo{author}{\bibfnamefont{P.~J.} \bibnamefont{Leek}},
  \bibinfo{author}{\bibfnamefont{A.}~\bibnamefont{Blais}}, \bibnamefont{and}
  \bibinfo{author}{\bibfnamefont{A.}~\bibnamefont{Wallraff}},
  \bibinfo{journal}{Nature} \textbf{\bibinfo{volume}{454}},
  \bibinfo{pages}{315} (\bibinfo{year}{2008}).

\bibitem[{\citenamefont{Blais et~al.}(2004)\citenamefont{Blais, Huang,
  Wallraff, Girvin, and Schoelkopf}}]{Blais2004}
\bibinfo{author}{\bibfnamefont{A.}~\bibnamefont{Blais}},
  \bibinfo{author}{\bibfnamefont{R.-S.} \bibnamefont{Huang}},
  \bibinfo{author}{\bibfnamefont{A.}~\bibnamefont{Wallraff}},
  \bibinfo{author}{\bibfnamefont{S.~M.} \bibnamefont{Girvin}},
  \bibnamefont{and} \bibinfo{author}{\bibfnamefont{R.~J.}
  \bibnamefont{Schoelkopf}}, \bibinfo{journal}{Phys. Rev. A}
  \textbf{\bibinfo{volume}{69}}, \bibinfo{pages}{062320}
  (\bibinfo{year}{2004}).

\bibitem[{\citenamefont{Schoelkopf and Girvin}(2008)}]{Schoelkopf2008}
\bibinfo{author}{\bibfnamefont{R.}~\bibnamefont{Schoelkopf}} \bibnamefont{and}
  \bibinfo{author}{\bibfnamefont{S.}~\bibnamefont{Girvin}},
  \bibinfo{journal}{Nature} \textbf{\bibinfo{volume}{451}},
  \bibinfo{pages}{664} (\bibinfo{year}{2008}).

\bibitem[{\citenamefont{Frunzio et~al.}(2005)\citenamefont{Frunzio, Wallraff,
  Schuster, Majer, and Schoelkopf}}]{Frunzio2005}
\bibinfo{author}{\bibfnamefont{L.}~\bibnamefont{Frunzio}},
  \bibinfo{author}{\bibfnamefont{A.}~\bibnamefont{Wallraff}},
  \bibinfo{author}{\bibfnamefont{D.}~\bibnamefont{Schuster}},
  \bibinfo{author}{\bibfnamefont{J.}~\bibnamefont{Majer}}, \bibnamefont{and}
  \bibinfo{author}{\bibfnamefont{R.}~\bibnamefont{Schoelkopf}},
  \bibinfo{journal}{IEEE Trans. Appl. Supercond.}
  \textbf{\bibinfo{volume}{15}}, \bibinfo{pages}{860} (\bibinfo{year}{2005}).

\bibitem[{\citenamefont{Baselmans et~al.}(2005)\citenamefont{Baselmans,
  Barends, Hovenier, Gao, Hoevers, de~Korte, and Klapwijk}}]{Baselmans2005}
\bibinfo{author}{\bibfnamefont{J.}~\bibnamefont{Baselmans}},
  \bibinfo{author}{\bibfnamefont{R.}~\bibnamefont{Barends}},
  \bibinfo{author}{\bibfnamefont{J.}~\bibnamefont{Hovenier}},
  \bibinfo{author}{\bibfnamefont{J.}~\bibnamefont{Gao}},
  \bibinfo{author}{\bibfnamefont{H.}~\bibnamefont{Hoevers}},
  \bibinfo{author}{\bibfnamefont{P.}~\bibnamefont{de~Korte}}, \bibnamefont{and}
  \bibinfo{author}{\bibfnamefont{T.}~\bibnamefont{Klapwijk}},
  \bibinfo{journal}{Bull. Soc. Roy. Sci. Li\`{e}ge}
  \textbf{\bibinfo{volume}{74}}, \bibinfo{pages}{5} (\bibinfo{year}{2005}).

\bibitem[{\citenamefont{Barends et~al.}(2007)\citenamefont{Barends, Baselmans,
  Hovenier, Gao, Yates, Klapwijk, and Hoevers}}]{Barends2007}
\bibinfo{author}{\bibfnamefont{R.}~\bibnamefont{Barends}},
  \bibinfo{author}{\bibfnamefont{J.}~\bibnamefont{Baselmans}},
  \bibinfo{author}{\bibfnamefont{J.}~\bibnamefont{Hovenier}},
  \bibinfo{author}{\bibfnamefont{J.}~\bibnamefont{Gao}},
  \bibinfo{author}{\bibfnamefont{S.}~\bibnamefont{Yates}},
  \bibinfo{author}{\bibfnamefont{T.}~\bibnamefont{Klapwijk}}, \bibnamefont{and}
  \bibinfo{author}{\bibfnamefont{H.}~\bibnamefont{Hoevers}},
  \bibinfo{journal}{IEEE Trans. Appl. Supercond.}
  \textbf{\bibinfo{volume}{17(2)}}, \bibinfo{pages}{263}
  (\bibinfo{year}{2007}).

\bibitem[{\citenamefont{O'Connell et~al.}(2008)\citenamefont{O'Connell,
  Ansmann, Bialczak, Hofheinz, Katz, Lucero, McKenney, Neeley, Wang, Weig
  et~al.}}]{Aaron2008}
\bibinfo{author}{\bibfnamefont{A.~D.} \bibnamefont{O'Connell}},
  \bibinfo{author}{\bibfnamefont{M.}~\bibnamefont{Ansmann}},
  \bibinfo{author}{\bibfnamefont{R.~C.} \bibnamefont{Bialczak}},
  \bibinfo{author}{\bibfnamefont{M.}~\bibnamefont{Hofheinz}},
  \bibinfo{author}{\bibfnamefont{N.}~\bibnamefont{Katz}},
  \bibinfo{author}{\bibfnamefont{E.}~\bibnamefont{Lucero}},
  \bibinfo{author}{\bibfnamefont{C.}~\bibnamefont{McKenney}},
  \bibinfo{author}{\bibfnamefont{M.}~\bibnamefont{Neeley}},
  \bibinfo{author}{\bibfnamefont{H.}~\bibnamefont{Wang}},
  \bibinfo{author}{\bibfnamefont{E.~M.} \bibnamefont{Weig}},
  \bibnamefont{et~al.}, \bibinfo{journal}{arXiv.org:0802.2404}
  (\bibinfo{year}{2008}).

\bibitem[{\citenamefont{Schreier et~al.}(2008)\citenamefont{Schreier, Houck,
  Koch, Schuster, Johnson, Chow, Gambetta, Majer, Frunzio, Devoret
  et~al.}}]{Schreier2008}
\bibinfo{author}{\bibfnamefont{J.~A.} \bibnamefont{Schreier}},
  \bibinfo{author}{\bibfnamefont{A.~A.} \bibnamefont{Houck}},
  \bibinfo{author}{\bibfnamefont{J.}~\bibnamefont{Koch}},
  \bibinfo{author}{\bibfnamefont{D.~I.} \bibnamefont{Schuster}},
  \bibinfo{author}{\bibfnamefont{B.~R.} \bibnamefont{Johnson}},
  \bibinfo{author}{\bibfnamefont{J.~M.} \bibnamefont{Chow}},
  \bibinfo{author}{\bibfnamefont{J.~M.} \bibnamefont{Gambetta}},
  \bibinfo{author}{\bibfnamefont{J.}~\bibnamefont{Majer}},
  \bibinfo{author}{\bibfnamefont{L.}~\bibnamefont{Frunzio}},
  \bibinfo{author}{\bibfnamefont{M.~H.} \bibnamefont{Devoret}},
  \bibnamefont{et~al.}, \bibinfo{journal}{Phys. Rev. B 77}
  \textbf{\bibinfo{volume}{77}}, \bibinfo{pages}{180502}
  (\bibinfo{year}{2008}).

\bibitem[{\citenamefont{Rabl et~al.}(2006)\citenamefont{Rabl, DeMille, Doyle,
  Lukin, Schoelkopf, and Zoller}}]{Rabl2006}
\bibinfo{author}{\bibfnamefont{P.}~\bibnamefont{Rabl}},
  \bibinfo{author}{\bibfnamefont{D.}~\bibnamefont{DeMille}},
  \bibinfo{author}{\bibfnamefont{J.~M.} \bibnamefont{Doyle}},
  \bibinfo{author}{\bibfnamefont{M.~D.} \bibnamefont{Lukin}},
  \bibinfo{author}{\bibfnamefont{R.~J.} \bibnamefont{Schoelkopf}},
  \bibnamefont{and} \bibinfo{author}{\bibfnamefont{P.}~\bibnamefont{Zoller}},
  \bibinfo{journal}{Phys. Rev. Lett.} \textbf{\bibinfo{volume}{97}},
  \bibinfo{pages}{033003} (\bibinfo{year}{2006}).

\bibitem[{Ver()}]{VeriCold}
\emph{\bibinfo{title}{VeriCold~{T}echnologies}},
  \urlprefix\url{http://www.vericold.com}.

\bibitem[{\citenamefont{Abdo et~al.}(2006)\citenamefont{Abdo, Segev,
  Shtempluck, and Buks}}]{Abdo2006}
\bibinfo{author}{\bibfnamefont{B.}~\bibnamefont{Abdo}},
  \bibinfo{author}{\bibfnamefont{E.}~\bibnamefont{Segev}},
  \bibinfo{author}{\bibfnamefont{O.}~\bibnamefont{Shtempluck}},
  \bibnamefont{and} \bibinfo{author}{\bibfnamefont{E.}~\bibnamefont{Buks}},
  \bibinfo{journal}{Phys. Rev. B} \textbf{\bibinfo{volume}{73}},
  \bibinfo{pages}{11} (\bibinfo{year}{2006}).

\bibitem[{\citenamefont{Martinis et~al.}(2005)\citenamefont{Martinis, Cooper,
  McDermott, Steffen, Ansmann, Osborn, Cicak, Oh, Pappas, Simmonds
  et~al.}}]{Martinis2005}
\bibinfo{author}{\bibfnamefont{J.~M.} \bibnamefont{Martinis}},
  \bibinfo{author}{\bibfnamefont{K.~B.} \bibnamefont{Cooper}},
  \bibinfo{author}{\bibfnamefont{R.}~\bibnamefont{McDermott}},
  \bibinfo{author}{\bibfnamefont{M.}~\bibnamefont{Steffen}},
  \bibinfo{author}{\bibfnamefont{M.}~\bibnamefont{Ansmann}},
  \bibinfo{author}{\bibfnamefont{K.~D.} \bibnamefont{Osborn}},
  \bibinfo{author}{\bibfnamefont{K.}~\bibnamefont{Cicak}},
  \bibinfo{author}{\bibfnamefont{S.}~\bibnamefont{Oh}},
  \bibinfo{author}{\bibfnamefont{D.~P.} \bibnamefont{Pappas}},
  \bibinfo{author}{\bibfnamefont{R.~W.} \bibnamefont{Simmonds}},
  \bibnamefont{et~al.}, \bibinfo{journal}{Phys. Rev. Lett.}
  \textbf{\bibinfo{volume}{95}}, \bibinfo{pages}{210503}
  (\bibinfo{year}{2005}).

\bibitem[{\citenamefont{Gevorgian et~al.}(1995)\citenamefont{Gevorgian,
  Linn\'{e}r, and Kollberg}}]{Gevorgian1995}
\bibinfo{author}{\bibfnamefont{S.}~\bibnamefont{Gevorgian}},
  \bibinfo{author}{\bibfnamefont{L.~J.~P.} \bibnamefont{Linn\'{e}r}},
  \bibnamefont{and} \bibinfo{author}{\bibfnamefont{E.~L.}
  \bibnamefont{Kollberg}}, \bibinfo{journal}{IEEE Trans. Microwave Theory
  Techn.} \textbf{\bibinfo{volume}{43(2)}}, \bibinfo{pages}{772}
  (\bibinfo{year}{1995}).

\bibitem[{\citenamefont{Watanabe et~al.}(1994)\citenamefont{Watanabe, Yoshida,
  Aoki, and Kohjiro}}]{Watanabe1994}
\bibinfo{author}{\bibfnamefont{K.}~\bibnamefont{Watanabe}},
  \bibinfo{author}{\bibfnamefont{K.}~\bibnamefont{Yoshida}},
  \bibinfo{author}{\bibfnamefont{T.}~\bibnamefont{Aoki}}, \bibnamefont{and}
  \bibinfo{author}{\bibfnamefont{S.}~\bibnamefont{Kohjiro}},
  \bibinfo{journal}{Jap. J. Appl. Phys.} \textbf{\bibinfo{volume}{33}},
  \bibinfo{pages}{5708} (\bibinfo{year}{1994}).

\bibitem[{\citenamefont{Chen and Chou}(1997)}]{Chen1997}
\bibinfo{author}{\bibfnamefont{E.}~\bibnamefont{Chen}} \bibnamefont{and}
  \bibinfo{author}{\bibfnamefont{S.}~\bibnamefont{Chou}},
  \bibinfo{journal}{IEEE Trans. Microwave Theory Techn.}
  \textbf{\bibinfo{volume}{45(6)}}, \bibinfo{pages}{939}
  (\bibinfo{year}{1997}).

\bibitem[{\citenamefont{Musil}(1986)}]{Musil1986}
\bibinfo{author}{\bibfnamefont{J.}~\bibnamefont{Musil}},
  \emph{\bibinfo{title}{Microwave Measurements of Complex Permittivity by Free
  Space Methods and Their Applications}} (\bibinfo{publisher}{Elsevier},
  \bibinfo{year}{1986}).

\bibitem[{\citenamefont{Tinkham}(1996)}]{Tinkham1996}
\bibinfo{author}{\bibfnamefont{M.}~\bibnamefont{Tinkham}},
  \emph{\bibinfo{title}{Introduction to Superconductivity}}
  (\bibinfo{publisher}{McGraw-Hill International Editions},
  \bibinfo{year}{1996}).

\bibitem[{\citenamefont{Parks}(1969)}]{Parks1969}
\bibinfo{author}{\bibfnamefont{R.}~\bibnamefont{Parks}},
  \emph{\bibinfo{title}{Superconductivity Vol.2}} (\bibinfo{publisher}{Marcel
  Dekker, Inc., New York}, \bibinfo{year}{1969}).

\bibitem[{\citenamefont{Poole}(1995)}]{Poole1995}
\bibinfo{author}{\bibfnamefont{C.~P.} \bibnamefont{Poole}},
  \emph{\bibinfo{title}{Superconductivity}} (\bibinfo{publisher}{Academic Press
  Inc.,U.S}, \bibinfo{year}{1995}).

\bibitem[{\citenamefont{Pozar}(1993)}]{Pozar1993}
\bibinfo{author}{\bibfnamefont{D.~M.} \bibnamefont{Pozar}},
  \emph{\bibinfo{title}{Microwave Engineering}}
  (\bibinfo{publisher}{Addison-Wesley Publishing Company},
  \bibinfo{year}{1993}).

\bibitem[{\citenamefont{Browne}(1987)}]{Browne1997}
\bibinfo{author}{\bibfnamefont{J.}~\bibnamefont{Browne}},
  \bibinfo{journal}{Microw. RF} \textbf{\bibinfo{volume}{26(2)}},
  \bibinfo{pages}{131} (\bibinfo{year}{1987}).

\end{thebibliography}
\end{document}